\documentclass{article}
\usepackage[utf8]{inputenc}
\usepackage{amsfonts}
\usepackage{amsmath}
\usepackage[table,xcdraw]{xcolor}
\usepackage{geometry}
\usepackage{multirow}
\usepackage{changepage}
\usepackage{xcolor}
\usepackage[ruled,vlined]{algorithm2e}
\usepackage{setspace} 
\usepackage[colorinlistoftodos]{todonotes}
\usepackage{color,soul}
\usepackage{url}

\def\StatVec{{T}}
\def\nwObs{{A^{\rm obs}}}
\def\nw{{A}}
\def\nwZero{{a}}
\def\nwSpace{{\mathcal{A}}}
\def\param{{\theta}}
\title{Improving ERGM Starting Values Using Simulated Annealing}

\date{July 2020}

\author{Christian S.~Schmid,
David R.~Hunter\\
Pennsylvania State University}
\usepackage{natbib}
\usepackage{graphicx}

\begin{document}

\maketitle

\maketitle
\doublespacing
\setstcolor{red}

\begin{abstract}
Much of the theory of estimation for exponential family models, which include exponential-family
random graph models (ERGMs) as a special case, is well-established and
maximum likelihood estimates in particular enjoy many desirable properties. 
However, in the case of many ERGMs, direct calculation of MLEs is impossible and therefore 
methods for approximating MLEs and/or alternative estimation methods must be employed.
Many MLE approximation methods require alternative estimates as starting points.  We discuss
one class of such alternatives here.
The MLE satisfies the so-called ``likelihood principle,'' unlike the MPLE.  This means that different
networks may have different MPLEs even if they have the same sufficient statistics.  We
exploit this fact here to search for improved starting values for approximation-based MLE methods.
The method we propose has shown its merit in producing an MLE for a network dataset and model that had
defied estimation using all other known methods.
\end{abstract}

\section{Maximum Likelihood Estimation for ERGMs}

Given a network-valued random variable $\nw$ and a set of sufficient statistics 
$\StatVec: \nwSpace \to \mathbb{R}^q ~, ~ \nwZero \mapsto (\StatVec_1(a) , \dots , \StatVec_q(a))$,
the exponential-family random graph model (ERGM) takes the form
\begin{equation}
    P_{\param}(\nw= \nwZero) = \frac{\exp \left\{ \param^\top  \StatVec(\nwZero) \right\}}{k(\param)}
    \qquad \mbox{for $\nwZero\in \nwSpace $,}
    \label{ergm}
\end{equation}
where $\nwSpace $ is the sample space of allowable networks and $\param \in \mathbb{R}^q$ is a vector of parameters. In many applications,
$\nwSpace $ denotes the entire set
$\{a \in \mathbb{R}^{N \times N}, a_{ij} = a_{ji}\in \{0,1\}, a_{ii}=0   \}$
of possible undirected networks on $N$ nodes, while
in other applications $\nwSpace $
may be constrained to be a proper subset
of this set.  (If we drop the requirement that $a_{ij}=a_{ji}$, then the sample space consists of directed networks.)
The sufficient statistics $\StatVec(\nwZero)$
play a central role in the model, since they enable the inclusion of traditional \textit{exogenous} covariates like a node's age as well as  
\textit{endogenous} statistics, i.e., statistics that allow for inference on the structure of the network. Popular endogenous statistics are a network's number of triangles or the number of pairs of ties that share one common node (two-stars). The normalizing constant 
\begin{equation}
k(\param) := \sum_{\nwZero^* \in \nwSpace } \exp \left\{\param^\top  T(\nwZero^*) \right\},
\end{equation} 
a weighted sum over all possible networks in the sample space,
assures that (\ref{ergm}) defines a probability model. 

Maximizing the log-likelihood function 
\begin{equation}
 \ell (\param) = \param^\top  \StatVec(\nwObs) - \log k(\param)
 \label{ML}
\end{equation}
of the ERGM results in the maximum likelihood estimator, or MLE.  The MLE can be difficult to calculate 
directly due to the 
required calculation of $k(\param)$, a sum which is only feasible for particular 
choices of $\StatVec(\nwZero)$ for which the sum may
be simplified or sample spaces small enough to allow direct calculation.  
Even for a small number of nodes, the sample space can be prohibitively large; 
for instance, there are over $3.5\times10^{13}$ networks on just $N=10$ nodes, 
and an additional node inflates 
this number by a factor of over 1000.
It is therefore often necessary to use an approximation method when an MLE
is desired, as exact calculation is not possible.

An alternative method of estimation, known
as maximum pseudolikelihood estimation (MPLE), has the desirable property that it is relatively easy to calculate even
in cases where an exact MLE is elusive.  This article first discusses approximate maximum likelihood estimation
and then summarizes how MPLE works and explains why, even in cases
where an MLE is desired, the approximation method often begins by calculating the MPLE.  Then, in Section~\ref{sec:SimAnn},
we explain a novel method to augment the approximation of the MLE by exploiting the 
failure of the MPLE to satisfy the so-called likelihood principle.  We demonstrate the use of this idea in Section~\ref{sec:Applications}.

\section{Approximate MLE and Exact MPLE}

An expedient to the problem of maximizing the often-intractable likelihood function~(\ref{ML})
is given by the Markov chain Monte Carlo maximum 
likelihood estimator (MCMLE).  First proposed by \cite{GeyerThompson1992} and then 
adapted to the ERGM framework by \cite{art:Snijders2002} and \cite{HunHan:06}.
The MCMLE is based on the idea that for any $\param_0 \in \mathbb{R}^q$, 
\begin{equation}\label{kRatio}
\frac{k(\param)}{k(\param_0)} = \mathbb{E}_{\param_0}
\exp\left\{ (\param - 
\param_0)^\top T(\nw) \right\},
\end{equation}
where the expectation is taken assuming that the random $\nw$ has the distribution $P_{\param_0}$.

For a given $\param_0$,
Equation~(\ref{kRatio}) may be exploited by 
sampling a large number of networks $A_1, \dots , A_L$ from the distribution $P_{\param_0}$.  Then one may 
approximate and optimize the difference of log-likelihood functions
\begin{equation}\label{loglikApprox}
\ell(\param) - \ell(\param_0) \approx (\param - \param_0)\cdot \StatVec(\nw) - \log 
\biggl( \frac{1}{L}\sum_{i=1}^L \exp \left\{ (\param - \param_0)^\top 
\StatVec(\nw_i)\right\} \biggr) .
\end{equation}

In theory, the MCMLE algorithm works for any starting value $\param_0$, however, 
\cite{HuHuHa:12} show that approximation~(\ref{loglikApprox}) is best when $\param$ is
close to $\param_0$, and furthermore the approximation can degrade badly when 
these parameter values are not close.  For this reason, in practice it is necessary to choose
$\param_0$ close to a maximizer of $\ell(\param)$ or else the MCMLE idea fails.  

The most common choice of $\param_0$ is the maximizer of the so-called pseudolikellhood function.
To construct the pseudolikelihood, we focus on the individual values of the tie indicators $\nw_{ij}$.
A straightforward calculation shows that for a particular $i, j$, the conditional distribution of $\nw_{ij}$ given
the rest of the network $\nw_{ij}^c$ is calculated from Equation~(\ref{ergm}) to be
\begin{equation}\label{dyad}
P_{\theta}(\nw_{ij}=1 | \nw_{ij}^c = \nwZero_{ij}^c) =  \mbox{logit}^{-1} \left[ \param^\top  
(\StatVec(\nwZero_{ij}^+) - \StatVec(\nwZero_{ij}^-) \right],
\end{equation}
where the inverse logit function is defined by $\mbox{logit}^{-1}(x) = \exp\{x\} / (1+\exp\{x\})$ and the networks 
$\nwZero_{ij}^+$ and $\nwZero_{ij}^-$ are formed by setting the value of $\nwZero_{ij}$ to be one or zero, respectively,
while fixing the rest of the network at $\nwZero_{ij}^c$.
We define $(\Delta \nwZero)_{ij} := \StatVec(\nwZero_{ij}^+) - \StatVec(\nwZero_{ij}^-)$ and refer to $(\Delta \nwZero)_{ij}$ as the $i,j$ vector of 
change statistics.  We may therefore rewrite Equation~(\ref{dyad}) as 
\[
P_{\theta}(\nw_{ij}=1 | \nw_{ij}^c = \nwZero_{ij}^c) =  \mbox{logit}^{-1} \left[ \param^\top  (\Delta \nwZero)_{ij} \right]
\]
for all $i,j$.  Under the additional assumption that the $\nw_{ij}$ are all independent of one another, these equations
together define a logistic regression model, and the maximum likelihood estimator for this logistic regression is known
as the maximum pseudo-likelihood estimator (MPLE) because the assumption of independence is not justified in all cases and therefore
the logistic regression likelihood function is sometimes misspecified.  Despite this misspecification, the MPLE is frequently used 
as an estimator of $\param$ because it is easily obtained using logistic regression software.  Indeed, the MPLE has a lengthy history
in the literature on ERGMs; see \citet{Schmid2020} for more details.

In particular, there is a substantial literature on the use of MPLE as an estimator in its own right.  For example, \citet{Schmid2020} argue that
when its covariance is properly estimated, the MPLE can allow for valid statistical inference just as the MLE can.  
On the other hand, for the most part it is assumed \citep[see, e.g.,][]{vanDuijn2009} that MLE is preferable to MPLE.  Indeed, one might
prefer MLE to MPLE simply based on the classical principle, as articulated by John Tukey for example, 
``Far better an approximate answer to the {\em right} question, which is often vague,
than an {\em exact} answer to the wrong question, which can always be made precise'' \citep{Tukey62}.
For the purpose of this article, the MPLE will serve merely as a value $\param_0$ used in Approximation~(\ref{loglikApprox})
and we assume that the ultimate goal is to obtain an approximate MLE via MCMLE.

\section{Comparison of MLE with MPLE}

The theory of exponential family models is well-established in general;
\citet{barndorff_nielsen78} gives an extensive book-length treatment of this theory.
As a particularly useful example in the context of ERGMs,
for exponential family distributions the 
expectation of the $\StatVec$ vector under the MLE equals the sufficient 
statistic on the observed network, i.e., $\mathbb{E}_{\hat{\param}}[\StatVec(\nw)] = 
\StatVec(\nwObs) $. In other words, the MLE equals the method of moments estimator. 
This aligns with one's general expectation that the distribution described by an 
estimate should on average describe the observed network.  Furthermore, this fact provides a useful
means for checking that a potential maximizer of the approximate log-likelihood function is in fact
close to the MLE, as one may simulate networks from the distribution derived from the MLE and check 
that their sample mean is approximately equal to $\StatVec(\nwObs)$.

\subsection{The Likelihood Principle}

Another property of the MLE is that it satisfies the likelihood 
principle \citep{Barnard:62,Birnbaum:62}, which states that all information in the data
relevant to the model parameters is contained in the 
likelihood function.  In the ERGM context, this means that an estimator should 
depend on $\nwObs$ only through $\StatVec(\nwObs)$, as the likelihood itself 
depends on $\nwObs$ only through $\StatVec(\nwObs)$.  
This means that two 
networks with the same sufficient statistic will yield the same MLE. 
However, as \cite{Corander1998} point out, the MPLE does not satisfy the 
likelihood principle.  This observation forms the basis of the remainder of this article.

\begin{figure}
\begin{center}
\includegraphics[scale=0.35]{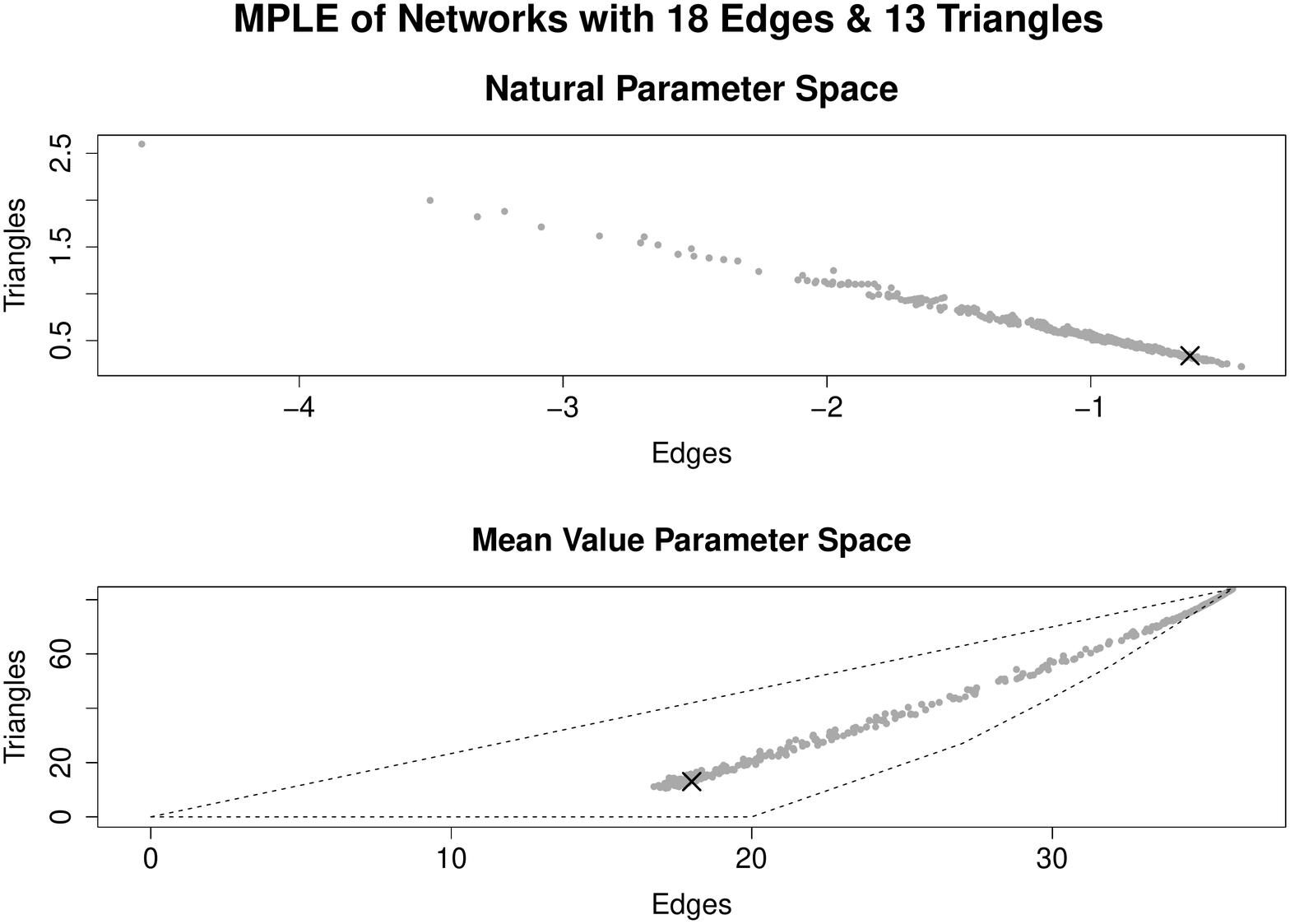}
\end{center}
\caption{Every grey dot represents the MPLE of a network on $N=9$ nodes with 18 
edges and 13 triangles using Model~\ref{ExampleERGM}.  The 'X' visualizes the MLE.}
\label{MPLE_18_13}
\end{figure}

The failure of the MPLE to satisfy the likelihood principle means that two networks $\nw^1$ and $\nw^2$
may have different MPLEs even if $\StatVec(\nw^1)=\StatVec(\nw^2)$.  We see this fact illustrated in 
Figure~ \ref{MPLE_18_13}, which depicts numerous possible MPLE values, each of which results from 
a network on 9 nodes with the same values of $\StatVec$.  
Also depicted in Figure~\ref{MPLE_18_13} is the mean value parameter space,
an alternative to the natural parameter space $\mathbb{R}^q$, 
which is defined as the interior of the sufficient statistic's sample space. 
Each parameter $\param$ can be uniquely projected to a point $g$ in mean value 
parameter space by a bijective function $\mu: \mathbb{R}^q \to \mathcal{C} $. 
For exponential family distributions, this function is defined as $\mu(\param) = 
\nabla \log k(\param) = \mathbb{E}_{\param}[T(Y)]$. In other words, a parameter 
$\param$'s corresponding point in mean value space is the expectation of the 
sufficient statistic with respect to the distribution defined by $\param$. This 
means that $\StatVec(\nwObs)$ is the MLE's projection into mean value parameter space,
which is an interesting fact in its own right, since the MLE in natural parameter space is 
hard to find, but finding its counterpart in mean value space is trivially easy.

More specifically, Figure~\ref{MPLE_18_13}
depicts the MLE and multiple MPLEs in the natural as well as in mean 
value parameter space of networks on $N=9$ nodes with exactly $18$ edges and 
$13$ triangles. The dashed lines in the lower plot visualize the boundary of the 
convex hull of the mean value parameter space. The networks were sampled from 
the $(\rho , \sigma , \tau)$-model of \citet{FrankStrauss:86} with $\sigma$ taken to be zero,
which results in an ERGM with a 2-dimensional $\StatVec$ vector.  We fixed $\rho = -1$
and $\tau =0.53 $, which is equivalent to 
\begin{equation}\label{ExampleERGM}
P_{\param}(\nw=\nwZero) \propto \exp \left\{-1 \cdot 
(\mbox{\# edges}) + 0.53 \cdot (\mbox{\# triangles}) \right\}.
\end{equation}
With only 9 nodes, it is computationally feasible to
enumerate all possible $T(\nwZero)$ vectors for $\nwZero\in\nwSpace$, so it is possible to calculate the MLE
exactly in this case.  \citet{Schmid2020} demonstrate how to achieve this enumeration using the 
{\tt ergm} package \citep{ergm} for the R computing environment \citep{r2020}.
Although it is possible to enumerate all network statistics according to their multiplicity in this
9-node example, it is not computationally feasible to calculate every possible MPLE resulting from a network
with 18 edges and 13 triangles.  Thus, Figure~\ref{MPLE_18_13} uses the simulated annealing method
explained in Section~\ref{sec:SimAnn} to generate multiple such networks randomly.

\subsection{Rao-Blackwellization}
Another potential way to exploit the failure of the MPLE to satisfy the likelihood principle is via the
so-called Rao-Blackwellization of the MPLE.  The Rao-Blackwell theorem \citep{lehmann1998} states that any 
estimator that is not a function of the sufficient statistic $\StatVec(\nwZero)$ is inadmissible, meaning that for any estimator 
$\Tilde{\param}$ that is not a function of $\StatVec(\nwZero)$, one can find an improved estimator $\param^*$ 
by taking the expectation of $\Tilde{\param}$ conditional on $\StatVec(\nwZero)$.  This new estimator has 
lower risk than $\Tilde{\param}$
by any convex lost function, e.g., mean squared error. 

In principle, computing the Rao-Blackwellized MPLE would require the distribution of the MPLE conditional on $\StatVec(\nwObs)$.
To put this idea into practice, 
a sample from this distribution could be obtained using, say, the simulated annealing method of Section~\ref{sec:SimAnn}.  However,
since the stochastic properties of the simulated annealing algorithm are not well understood, we have not explored this particular
idea in the current manuscript.

\section{New Starting Values Via Simulated Annealing}
\label{sec:SimAnn}

As described earlier, maximum likelihood estimation by MCMC requires the 
selection of an auxiliary parameter $\param_0$. Even though in theory the 
algorithm operates with any choice of $\param_0$, in practice, a parameter close 
to the MLE is essential for the MCMLE algorithm to work successfully 
\citep{HuHuHa:12}. The commonly used starting value for $\param_0$ is the MPLE, since it is 
simple to calculate via logistic regression; even though other methods have been proposed 
\citep[e.g.,][]{Kriv:17}, MPLE remains the predominant method. As shown in 
Figure~\ref{MPLE_18_13}, however, the MPLE can in some cases be very different from the MLE, 
which typically results in the failure of the MCMLE algorithm.
In our 9-node example, the MCMLE algorithm failed 
for about a third of the MPLEs as starting values shown in Figure~\ref{MPLE_18_13}.

\subsection{Failure of the MCMLE algorithm:  An Illustrative Example} 

\begin{figure}
\vspace*{-15mm}
\begin{center}
\includegraphics[scale=0.35]{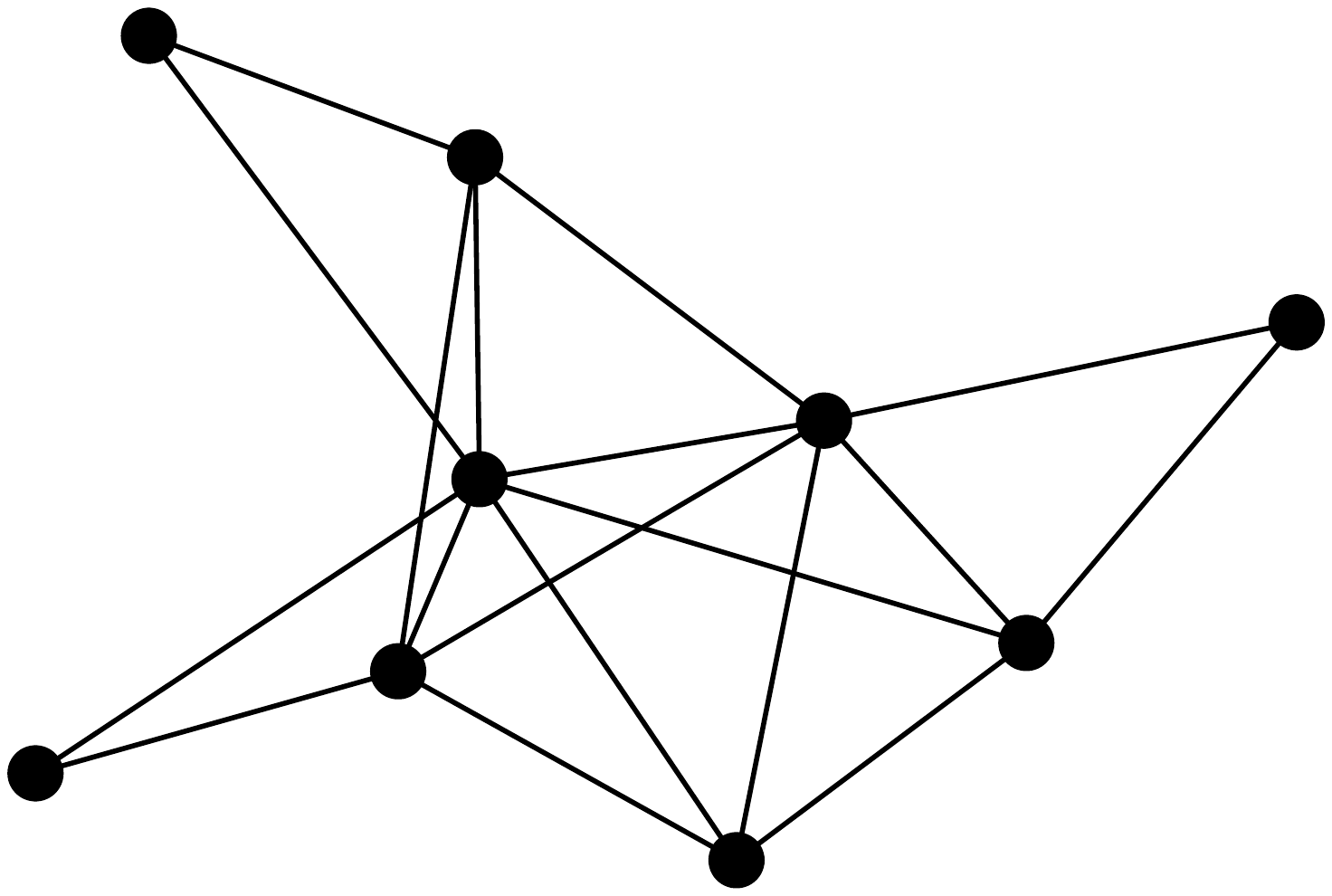}
\end{center}
\vspace*{-15mm}
\caption{A network with 18 edges and 13 triangles.}
\label{network16}
\end{figure}

\begin{figure}
\begin{center}
\includegraphics[scale=0.32]{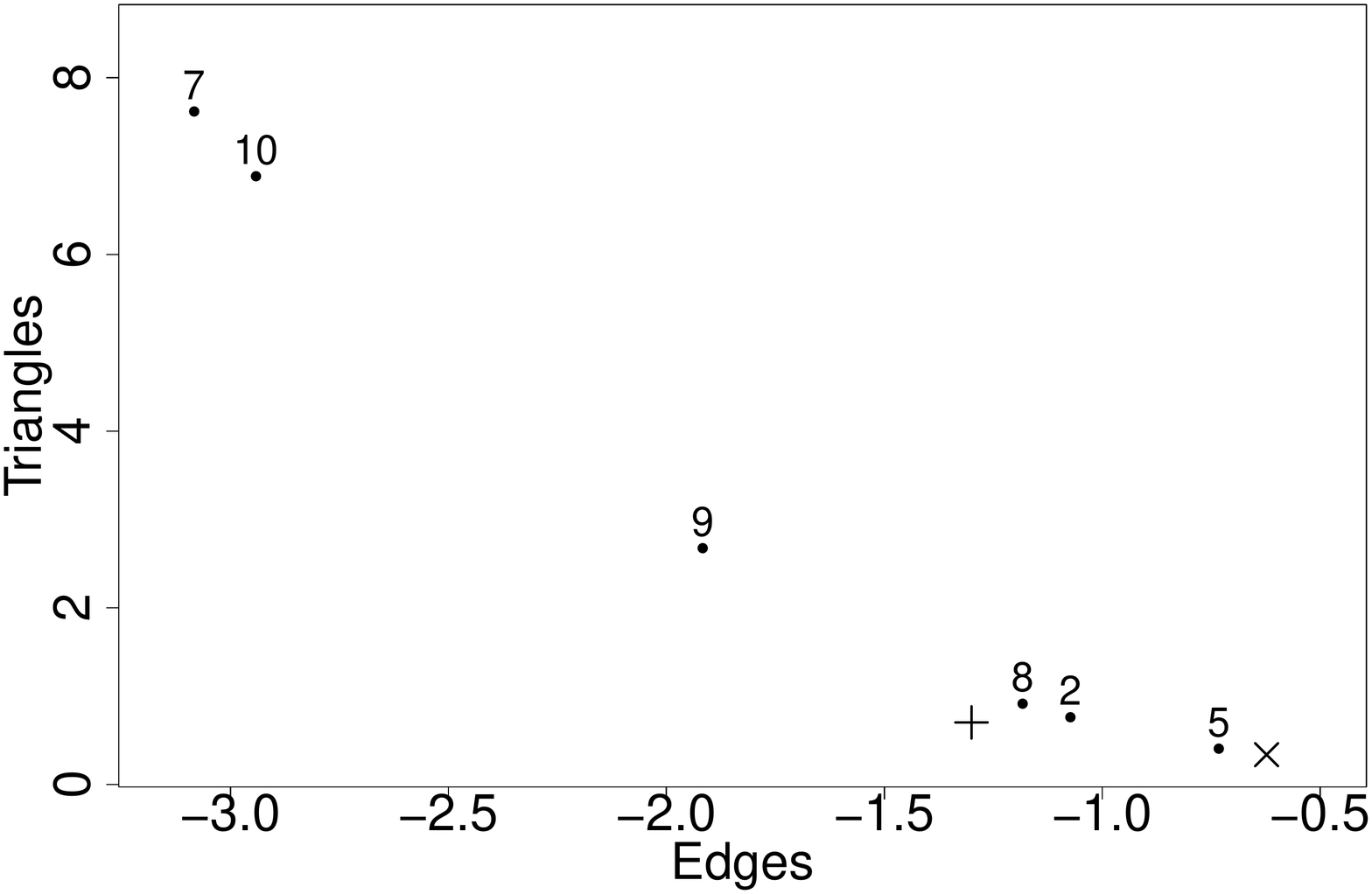}
\end{center}
\caption{MCMLEs of the network in Figure \ref{network16} using Model~(\ref{ExampleERGM})
for ten independent trials. '+' indicates the MPLE, 'x' the MLE.}
\label{differentMCMLEs}
\end{figure}

We demonstrate the problems of the MPLE as starting value for the MCMLE 
algorithm by taking a particular network on $9$ nodes with $18$ edges and $13$ 
triangles as depicted in Figure \ref{network16}. The MPLE of this particular 
network is $\Tilde{\param} = (-1.3 , 0.702)$, while the MLE  
can be exactly calculated as $\hat{\param} = (-0.623 , 0.337)$. 
Table~\ref{differentMCMLEs} shows MCMLE results for the same network, where the 
algorithm was initialized by different $\Tilde{\param}$ values at each trial from among the
values depicted in Figure~\ref{MPLE_18_13}. 
In four out of ten trials, the algorithm stopped due to model degeneracy. 
As first sketched by \cite{Handcock2001} and later studied in detail by
\cite{Schweinberger2011}, degeneracy occurs when the distribution defined by 
$P_{\hat\param}$---ostensibly the best possible parameter value, in some sense---places 
most of its probability mass on just a few networks, usually the 
empty and the full network. In this scenario, the simulated networks are so 
different from the observed network that the MCMLE algorithm fails. In six cases
shown in Table~\ref{differentMCMLEs}, 
the algorithm provided an MCMLE. Yet three trials yielded
estimates further away from the 
MLE than the MPLE, leading to estimates in natural 
parameter space close to the boundary of the convex hull and therefore clearly 
different from the true MLE based on the observed network. The only glimmer of hope is 
trial 5, which leads to an estimate somewhat similar to the MLE.  
This example shows that it is necessary to check that an MLE is producing values of
$\StatVec$, on average, close to $\StatVec(\nwObs)$.  Once this takes place, additional
techniques such as those detailed in \citet{HuHuHa:12} may be employed to fine-tune 
the MLE.  However, even using such techniques,
many of the trials from Table~\ref{differentMCMLEs} did not end successfully.

\subsection{Simulated Annealing and MPLE}

As noted earlier, 
\cite{HuHuHa:12} show that a poorly chosen $\param_0$ can make successful MCML 
estimation almost impossible.
Consequently, the more different the MPLE is from the MLE, the more difficult it 
is to find the MCMLE, since the MPLE is commonly taken as the algorithm's 
starting value $\param_0$ due to its simple and fast calculation. Inspired by 
Figure \ref{MPLE_18_13}, 
we propose a novel approach for finding an improved starting value for the MCMLE 
algorithm, namely,  
to search for networks that yield the same---or at least nearly the same---
statistics as the observed network, then consider these networks' MPLEs as 
potential starting values.

We propose searching for networks with the same statistics as the observed 
network using the \textit{simulated annealing} algorithm \citep{Kirkpatrick83}. 
This algorithm was initially inspired from the practice of annealing metal, a procedure
in which the material is heated and then cooled very slowly, allowing it to be 
gradually shaped into the desired form. The process of heating up and cooling 
down is translated into the simulated annealing approach by allowing interim 
results initially (during the ``heated'' phase) to be worse than previous results; that is, 
the algorithm is more stochastic than deterministic in its early phase. Gradually, the
algorithm ``cools,'' becoming more deterministic and less prone to random jumps. The 
hope is that the algorithm does not get stuck at local maxima that it otherwise 
might not be able to leave if a purely deterministic optimization algorithm were used. 
By starting in an initial random phase followed by increasingly 
deterministic behavior, the algorithm can find and then focus on areas of the search
space that include globally optimal solutions.

\section{Applications}
\label{sec:Applications}

We demonstrate the simulated annealing-based approach to MCMLE
using the E.~coli transcriptional regulation 
network of \cite{Shen-Orr.2002}, which is based on the RegulonDB data of 
\cite{Salgado.2001}. The nodes in this network represent operons, while an edge 
from operon $i$ to $j$ indicates that $i$ encodes a transcription factor that 
regulates $j$. Even though this is originally a directed network that contains 
self-edges, i.e., operons that regulate themselves, we follow 
\cite{SaulFilkov2007} and treat this network as undirected and without 
self-edges. This results in a network with 519 edges and 418 nodes. We study the 
same ERGM on these data as \cite{HuHuHa:12}, a model which yields MCMLEs 
considerably different from the MPLE, making estimation difficult. The model's 
statistics consist of the number of edges; the numbers of nodes with degrees two, 
three, four, and six; and the geometrically weighted degree distribution with the 
decay parameter fixed at $0.25$. As demonstrated by \cite{HuHuHa:12}, 
initializing the MCMLE algorithm with the MPLE does not produce successful 
results.

\begin{figure}
\begin{center}
\includegraphics[scale=0.45]{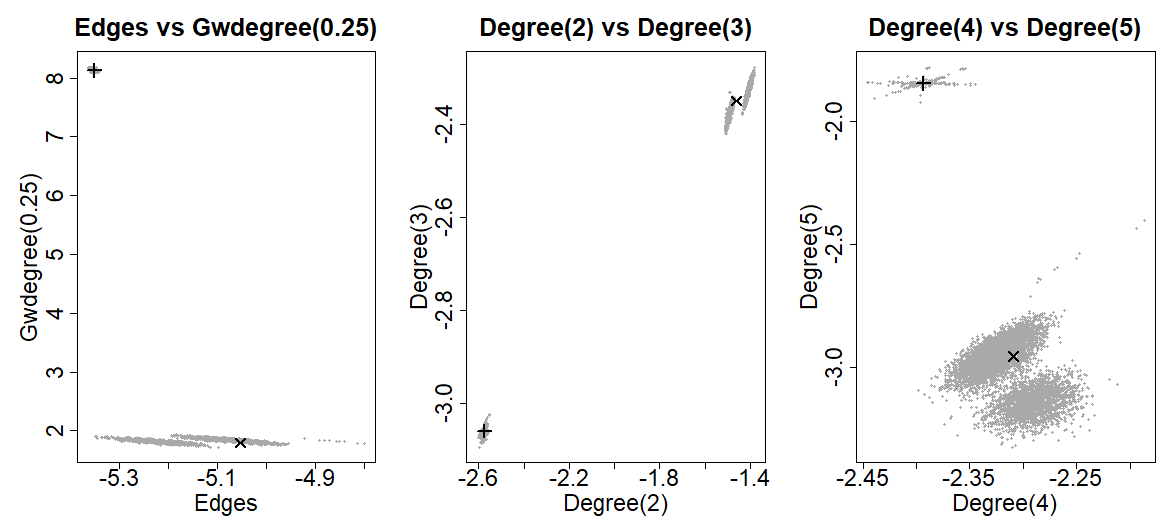}
\end{center}
\vspace*{-4mm}
\caption{MPLEs of networks with the same sufficient statistics as the observed 
E.~coli network. The MPLE of the observed network is marked with an '+', the 
MCMLE with an 'x'.}
\label{SAN}
\end{figure}

Figure \ref{SAN} depicts the MPLEs of 10,000 networks (in grey) that have the 
same sufficient statistics as the original network, including the MPLE of the original 
network as well as the MCMLE.  All $10,000$ 
networks were obtained using the simulated annealing algorithm, always 
starting with the observed network. It is remarkable that the MPLEs of these 
networks essentially form two clusters, one cluster that includes the MPLE and 
another one that includes the MCMLE of the observed network. Among other things, this 
figure illustrates why finding the MCMLE by beginning the algorithm at the 
observed network's MPLE is a difficult task. The MPLE is simply not close enough to the MLE
for the approximate log-likelihood using an MPLE-based sample of networks to 
be effective.  On the other 
hand, setting the starting value to one of the MPLEs that form the cluster 
around the MCMLE evidently makes the algorithm more likely to succeed.

\begin{figure}
\hspace*{-15mm} 
\includegraphics[scale=0.6]{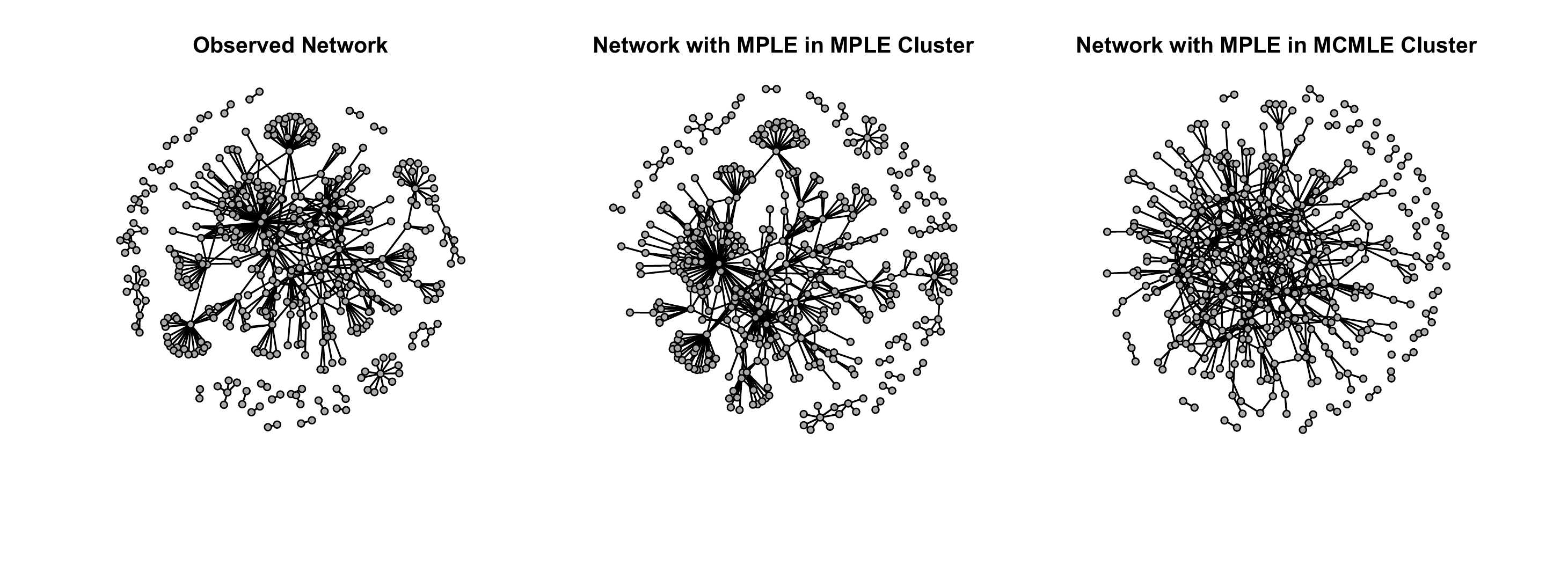}
\vspace*{-15mm}
\caption{The left visualizes the E.coli transcriptional regulation
network of \cite{Shen-Orr.2002}. The center depicts a network whose MPLE falls 
into the MPLE cluster of Figure \ref{SAN}. The right depicts a network whose 
MPLE falls into the MPLE cluster of Figure \ref{SAN}. All three networks have 
the same sufficient statistics vector. } 
\label{Ecoli_plots}
\end{figure}

A natural question that arises when considering Figure \ref{SAN} is how the 
networks in the two clusters differ. Figure \ref{Ecoli_plots} visualizes 
networks of each cluster as well as the original E.~coli network. Even though 
all three networks have the same sufficient statistics, it is possible to discern that 
the network in the MPLE cluster maintains some of the distinctive structures of 
the original network, while the network in the MCMLE cluster is more reminiscent 
of an Erd\"os-R\'enyi network, i.e., a network where each tie occurs independently 
with same probability $p$. The fact that the occurrence of a tie in an 
Erd\"os-R\'enyi network does not depend on the occurrence or absence of any other 
tie results in a model in which the MPLE and the MLE are the same. 
Consequently, we conjecture that it is advantageous to find a 
network that resembles an Erd\"os-R\'enyi network and that, in addition, yields the same 
sufficient statistics as the observed network.

Simulating networks from the probability distribution defined by the obtained 
MCMLE results in networks that resemble the rightmost network in Figure~\ref{Ecoli_plots}, 
rather than the observed network. This consequently casts 
doubt on whether the statistics of this ERGM appropriately capture the unique structure 
of the observed network. Stated differently, if the MLE, ostensibly the gold standard among estimators,
yields unsatisfying results, the model itself should be reconsidered. 
It is important to remember that in such cases, as indeed in this case,
generally the MPLE also fails to result
in simulated networks resembling the original network; thus, a poor model cannot
generally be mended by using a different estimation technique.

We repeat the simulation study that resulted in Figure~\ref{SAN}, with the exception that
we start the simulated annealing algorithm with an Erd\"os-R\'enyi-generated 
network with similar density to the original network, i.e., where $p$ is defined as 
the ratio of the number of ties to the number of possible ties. With this modification,
the simulated annealing algorithm only finds networks 
where the MPLE is in the proximity of the MCMLE, meaning that the MPLE of any of 
these networks is an improved starting value for the MCMLE algorithm compared to 
the original MPLE.

In summary, we suggest finding an improved starting value $\param_0$ the 
following way:

\begin{enumerate}
    \item For a given network $\nwObs$, find the sufficient statistics $\StatVec(\nwObs)$
    \item Simulate a network $G$ using an Erd\"os-R\'enyi model on the same nodes as $\nwObs$ 
    and with $p$ given by the edge density of $\nwObs$
    \item Simulate a new network $\nw^*$ with $\StatVec(\nw^{obs})= \StatVec(\nw^*)$ with simulated 
    annealing and start with the algorithm at $G$. If $\StatVec(\nw)$ includes continuous 
    statistics, find a network $A^*$ with $\StatVec(\nw) \approx \StatVec(\nw^*)$ 
    \item Take the MPLE of $A^*$ as the improved starting value $\param_0$
\end{enumerate}

This approach was successfully applied by \cite{SchmidChenDesmarais} to a 
citation network based on all US Supreme Court 
majority opinions written between 1937 and 2015. The majority opinion of each 
court case is considered a node, and a citation from case $i$ to case $j$ is defined as 
a directed tie. The network includes $10{,}020$ nodes and the ERGM used includes $14$ 
endogenous and exogenous statistics---complicated enough to be computationally 
challenging.  Ultimately, \cite{SchmidChenDesmarais} find that
multiple different approaches fail to yield a successful MCMLE, among them
the stepping method of 
\citet{HuHuHa:12} and the standard MCMLE algorithm using the MPLE as starting value. 
Yet the approach described here---which reproduces the observed network statistics only
approximately due to the inclusion of several continuous variables among these
statistics---does produce a successful MCMLE; it is the only known method that does so in this example.

\section{Discussion}

The basic idea of this article---searching for networks that have the same, or approximately the same, ERGM statistics as the
observed network and then using the MPLEs from these networks as starting values for a traditional MCMLE algorithm---is
relatively simplistic.  Yet it has already demonstrated its value in solving previously unsolvable ERGM-based estimation problems.

It seems there is still much to learn about this methodology.  For instance, is there a way to implement Rao-Blackwellization, and does
this approach lead to estimates that are closer in their behavior to the MLE?  
Also, how important is it to search the sample space of all networks 
thoroughly, and is an alternative to simulated annealing possible for this purpose?  Figure~\ref{SAN} suggests that initializing the simulated 
annealing algorithm at the observed network and the Erd\"os-R\'enyi-generated network generate wholly distinct clouds of MPLE values, which
leads to the question of whether {\em all} possible MPLE values for a particular set of statistics is actually bimodal, or whether in fact there is
a vast as-yet-unexplored set of MPLE values that could potentially be of use both as starting MCMLE values and as sample points in a
Rao-Blackwellization scheme.  

That such a simple idea can prove so effective relative to all other known methods suggests that there exists immense untapped potential 
for improving upon approximate likelihood-based inference using MPLEs.

\addcontentsline{toc}{section}{Bibliography}
\bibliography{references} 

\begin{thebibliography}{}

\bibitem[\protect\citeauthoryear{Barnard, Jenkins, and Winsten}{Barnard
  et~al.}{1962}]{Barnard:62}
Barnard, G.~A., G.~M. Jenkins, and C.~B. Winsten (1962).
\newblock Likelihood inference and time series.
\newblock {\em Journal of the Royal Statistical Society. Series A
  (General)\/}~{\em 125\/}(3), 321--372.

\bibitem[\protect\citeauthoryear{Barndorff-Nielsen}{Barndorff-Nielsen}{1978}]{barndorff_nielsen78}
Barndorff-Nielsen, O. (1978).
\newblock {\em Information and Exponential Families in Statistical Theory}.
\newblock Wiley.

\bibitem[\protect\citeauthoryear{Birnbaum}{Birnbaum}{1962}]{Birnbaum:62}
Birnbaum, A. (1962).
\newblock On the foundations of statistical inference.
\newblock {\em Journal of the American Statistical Association\/}~{\em
  57\/}(298), 269--306.

\bibitem[\protect\citeauthoryear{Corander, Dahmstr\"om, and
  Dahmstr\"om}{Corander et~al.}{1998}]{Corander1998}
Corander, J., K.~Dahmstr\"om, and P.~Dahmstr\"om (1998).
\newblock Maximum likelihood estimation for {M}arkov graphs.
\newblock Technical report, Department of Statistics, University of Stockholm.

\bibitem[\protect\citeauthoryear{Frank and Strauss}{Frank and
  Strauss}{1986}]{FrankStrauss:86}
Frank, O. and D.~Strauss (1986).
\newblock Markov graphs.
\newblock {\em Journal of the American Statistical Association\/}~{\em
  81\/}(395), 832--842.

\bibitem[\protect\citeauthoryear{Geyer and Thompson}{Geyer and
  Thompson}{1992}]{GeyerThompson1992}
Geyer, C.~J. and E.~A. Thompson (1992).
\newblock Constrained {M}onte {C}arlo maximum likelihood for dependent data.
\newblock {\em Journal of the Royal Statistical Society. Series B
  (Methodological)\/}~{\em 54\/}(3), 657--699.

\bibitem[\protect\citeauthoryear{Handcock}{Handcock}{2001}]{Handcock2001}
Handcock, M.~S. (2001).
\newblock Assessing degeneracy in statistical models of social networks.
\newblock Technical report, Center for Statistics and the Social Sciences,
  University of Washington.

\bibitem[\protect\citeauthoryear{Handcock, Hunter, Butts, Goodreau, Krivitsky,
  and Morris}{Handcock et~al.}{2019}]{ergm}
Handcock, M.~S., D.~R. Hunter, C.~T. Butts, S.~M. Goodreau, P.~N. Krivitsky,
  and M.~Morris (2019).
\newblock {\em ergm: Fit, Simulate and Diagnose Exponential-Family Models for
  Networks}.
\newblock The Statnet Project (\url{https://statnet.org}).
\newblock R package version 3.10.0-4837.

\bibitem[\protect\citeauthoryear{Hummel, Hunter, and Handcock}{Hummel
  et~al.}{2012}]{HuHuHa:12}
Hummel, R.~M., D.~R. Hunter, and M.~S. Handcock (2012).
\newblock Improving simulation-based algorithms for fitting {ERGM}s.
\newblock {\em Journal of Computational and Graphical Statistics\/}~{\em
  21\/}(4), 920--939.

\bibitem[\protect\citeauthoryear{Hunter and Handcock}{Hunter and
  Handcock}{2006}]{HunHan:06}
Hunter, D.~R. and M.~S. Handcock (2006).
\newblock Inference in curved exponential family models for networks.
\newblock {\em Journal of Computational and Graphical Statistics\/}~{\em
  15\/}(3), 565--583.

\bibitem[\protect\citeauthoryear{Kirkpatrick, Gelatt, and Vecchi}{Kirkpatrick
  et~al.}{1983}]{Kirkpatrick83}
Kirkpatrick, S., C.~D. Gelatt, and M.~P. Vecchi (1983).
\newblock Optimization by simulated annealing.
\newblock {\em Science\/}~{\em 220\/}(4598), 671--680.

\bibitem[\protect\citeauthoryear{Krivitsky}{Krivitsky}{2017}]{Kriv:17}
Krivitsky, P.~N. (2017).
\newblock Using contrastive divergence to seed {M}onte {C}arlo {MLE} for
  exponential-family random graph models.
\newblock {\em Computational Statistics and Data Analysis\/}~{\em 107\/}(C),
  149--161.

\bibitem[\protect\citeauthoryear{Lehmann and Casella}{Lehmann and
  Casella}{1998}]{lehmann1998}
Lehmann, E. and G.~Casella (1998).
\newblock {\em Theory of Point Estimation}.
\newblock Springer Verlag.

\bibitem[\protect\citeauthoryear{{R Core Team}}{{R Core Team}}{2020}]{r2020}
{R Core Team} (2020).
\newblock {\em R: A Language and Environment for Statistical Computing}.
\newblock Vienna, Austria: R Foundation for Statistical Computing.

\bibitem[\protect\citeauthoryear{Salgado, Santos-Zavaleta, Gama-Castro,
  Mill{\'a}n-Z{\'a}rate, D{\'\i}az-Peredo, S{\'a}nchez-Solano, P{\'e}rez-Rueda,
  Bonavides-Mart{\'\i}nez, and Collado-Vides}{Salgado
  et~al.}{2001}]{Salgado.2001}
Salgado, H., A.~Santos-Zavaleta, S.~Gama-Castro, D.~Mill{\'a}n-Z{\'a}rate,
  E.~D{\'\i}az-Peredo, F.~S{\'a}nchez-Solano, E.~P{\'e}rez-Rueda,
  C.~Bonavides-Mart{\'\i}nez, and J.~Collado-Vides (2001).
\newblock Regulon{DB} (version 3.2): Transcriptional regulation and operon
  organization in {E}scherichia coli {K}-12.
\newblock {\em Nucleic acids research\/}~{\em 29\/}(1), 72--74.

\bibitem[\protect\citeauthoryear{Saul and Filkov}{Saul and
  Filkov}{2007}]{SaulFilkov2007}
Saul, Z. and V.~Filkov (2007).
\newblock Exploring biological network structure using exponential random graph
  models.
\newblock {\em Bioinformatics (Oxford, England)\/}~{\em 23}, 2604--11.

\bibitem[\protect\citeauthoryear{Schmid, Chen, and Desmarais}{Schmid
  et~al.}{2020}]{SchmidChenDesmarais}
Schmid, C.~S., T.~H. Chen, and B.~A. Desmarais (2020).
\newblock Generative dynamics of {S}upreme {C}ourt citations.
\newblock {\em Working Paper, The Pennsylvania State University\/}.

\bibitem[\protect\citeauthoryear{Schmid and Hunter}{Schmid and
  Hunter}{2020}]{Schmid2020}
Schmid, C.~S. and D.~R. Hunter (2020).
\newblock Accounting for model misspecification when using pseudolikelihood for
  {ERGM}s.
\newblock Unpublished manuscript.

\bibitem[\protect\citeauthoryear{Schweinberger}{Schweinberger}{2011}]{Schweinberger2011}
Schweinberger, M. (2011).
\newblock Instability, sensitivity, and degeneracy of discrete exponential
  families.
\newblock {\em Journal of the American Statistical Association\/}~{\em
  106\/}(496), 1361--1370.

\bibitem[\protect\citeauthoryear{Shen-Orr, Milo, Mangan, and Alon}{Shen-Orr
  et~al.}{2002}]{Shen-Orr.2002}
Shen-Orr, S., R.~Milo, S.~Mangan, and U.~Alon (2002, 06).
\newblock Network motifs in the transcriptional regulation network of
  {E}scherichia coli.
\newblock {\em Nature genetics\/}~{\em 31}, 64--68.

\bibitem[\protect\citeauthoryear{Snijders}{Snijders}{2002}]{art:Snijders2002}
Snijders, T.~A. (2002).
\newblock Markov chain {M}onte {C}arlo estimation of exponential random graph
  models.
\newblock {\em Journal of Social Structure\/}~{\em 3\/}(2), 1--40.

\bibitem[\protect\citeauthoryear{Tukey}{Tukey}{1962}]{Tukey62}
Tukey, J.~W. (1962).
\newblock The future of data analysis.
\newblock {\em Annals of Mathematical Statistics\/}~{\em 33}, 1--67.

\bibitem[\protect\citeauthoryear{van Duijn, Gile, and Handcock}{van Duijn
  et~al.}{2009}]{vanDuijn2009}
van Duijn, M.~A., K.~J. Gile, and M.~S. Handcock (2009).
\newblock A framework for the comparison of maxmimum pseudo-likelikood and
  maximum likelihood estimation of exponential family random graph models.
\newblock {\em Social Networks\/}~{\em 31\/}(1), 52--62.

\end{thebibliography}
\bibliographystyle{chicago}

\end{document}